\documentclass{ws-mpla}
\usepackage{graphics}
\usepackage{epsfig}
\usepackage[latin1]{inputenc}
\usepackage{cite}

\newcommand{\dg}{\ensuremath{^{\circ}}}
\def\xbj{\ensuremath{x_{Bj}}}
\newcommand{\gev}{\,\mbox{GeV}}

\newcommand{\gevsq}{\ensuremath{\mathrm{GeV}^2} }

\def\prp{\perp}
\def\prp{t}

\def\sx{small-$x$}
\def\kt{\ensuremath{k_\prp}}
\def\kti#1{\ensuremath{k_{\prp #1}}}
\newcommand{\ktp}{k_{\prp}^{\prime}}
\def\pt{\ensuremath{p_\prp}}
\def\pti#1{\ensuremath{p_{\prp #1}}}

\def\pythia{{\sc Pythia}}
\def\cascade{{\sc Cascade}}
\def\rapgap{{\sc Rapgap}}

\def\herwig{{\sc Herwig}}
\def\ldcmc{{\sc Ldc}}

\def\as{\ensuremath{\alpha_s}}
\def\sub#1{\ensuremath{_{\mbox{\scriptsize #1}}}}
\def\alb{\ensuremath{\bar{\alpha}\sub{s}}}

\newcommand{\qb}{\ensuremath{\bar{q}} }
\newcommand{\Pmax}{\bar{q}}

\newcommand{\CCFM}{CCFMa,CCFMb,CCFMc,CCFMd}
\newcommand{\BFKL}{BFKLa,BFKLb,BFKLc}
\newcommand{\LDCMC}{LDCa,LDCb,LDCc,LDCd}
\newcommand{\alphasb}{\alb}

\newcommand{\PYTHIAMC}{pythia61}
\newcommand{\RAPGAPMC}{RAPGAP206}
\newcommand{\HERWIGMC}{Herwig}
\newcommand{\DGLAP}{DGLAPa,DGLAPb,DGLAPc,DGLAPd}
\newcommand{\asb}{\alb}

\newcommand{\CASCADEMC}{jung_salam_2000,CASCADEMC}

\newcommand{\hqhera}{H1_f2charm_2000,Baranov:2002,Jung-ascona,jung-hq-2001}
\newcommand{\jetshera}{H1fwdjet2,jung_salam_2000,H1dijet,kamil}

\begin{document}

\markboth{Hannes Jung}{}

%
\catchline{}{}{}{}{}
%

\title{\kt - factorization and CCFM\\
the solution for describing the hadronic final states - everywhere ?
}

\author{Hannes Jung}

\address{Department of Physics, Lund University, Box 118\\
Lund, 22100\\
Sweden\\
hannes.jung@desy.de}

\maketitle

\pub{Received (Day Month Year)}{Revised (Day Month Year)}

\begin{abstract}
The basic ideas of \kt-factorization and CCFM parton evolution is discussed. The
unintegrated gluon densities, obtained from CCFM fits to the proton structure
function data at HERA are used to predict hadronic final state cross sections
like jet production at HERA, but also comparisons with recent measurements of
heavy quark production at the Tevatron are presented. Finally,  the
\kt-factorization approach is applied to Higgs production at high energy hadron
hadron colliders and 
the transverse
momentum spectrum of Higgs production at the LHC is calculated.
\par
This paper is dedicated to the memory of 
Bo Andersson and Jan Kwiecinski, who inspired so much in the field of \sx\ physics
and passed away too early. 
\keywords{Keyword1; keyword2; keyword3.}
\end{abstract}

\ccode{PACS Nos.: include PACS Nos.}

\section{Introduction}
The theory of strong interaction, QCD, has been very successful in describing
many experimental measurements, but a number of problems have not yet been
solved, some of which are related to the transition between the 
perturbative and  non-perturbative region. The 
perturbative methods, however, work surprisingly well, 
even down to very low scales, where the running coupling
constant, $\as$, starts to become large. 
Another type of problem is related to the observation, that at high energies,
even for small values of coupling constant $\as$, the phase space for parton  
emissions increases fast, and that therefore it is not sufficient to
include only a few calculable terms in the perturbative expansion. 
This problem can be treated by resumming the leading logarithmic behavior of the
cross section. The most important contribution 
at small $x$ is gluon bremsstrahlung,
with the typical behavior of being largest in the  infrared and/or  collinear 
region. Two different resummation strategies have
been developed: the DGLAP~\cite{\DGLAP} approach, resumming leading logarithms
of ratios of subsequent virtualities also called collinear
approach, and the BFKL~\cite{\BFKL} approach, resumming the infrared
contributions, also called the \kt-factorization~\cite{CCH,CE} or the semi-hard
approach~\cite{GLR,LRSS2}.
The CCFM~\cite{\CCFM} approach attempts to cover both the collinear and the
infrared regions by considering color coherence effects, and in the limit of
asymptotic energies is almost
equivalent~\cite{Forshaw:1998uq,Webber:1998we,Salam:1999ft} to the 
BFKL and 
DGLAP evolution equations. The LDC~
approach~\cite{\LDCMC}, which is a reformulation of CCFM, 
a unified DGLAP-BFKL 
approach~\cite{Martin_Stasto,Martin_Sutton,martin_kimber} and the approach of
doubly unintegrated parton distributions~\cite{watt-martin-ryskin} 
are not further discussed in this report.
\par
In the collinear approach 
the cross section is factorized into a process dependent hard
scattering matrix element convoluted with universal parton density functions.
Since strong ordering in virtualities is required in the evolution,
the largest virtuality is in the hard scattering and therefore the
virtuality of the partons entering the hard scattering matrix element can be
neglected and they can be treated as being
collinear with the incoming 
hadron\footnote{However neglecting the transverse momentum of the partons even in the
collinear approach has been criticized in Ref.~\refcite{Collins-collfac} 
as being
unnecessary and unphysical.}.
Any physics process in the fixed order 
collinear factorization scheme is then calculated by a convolution of a process
dependent coefficient function
$C^a(\frac{x}{z})$ with collinear (independent of $k_t$) 
parton density functions at a scale $\mu_f^2$ 
(e.g. $\mu_f^2=Q^2$ in $F_2(x,Q^2)$
in deep inelastic scattering (DIS)):
\begin{equation}
\sigma = \sigma_0 \int \frac{dz}{z} C^a(\frac{x}{z}) f_a(z,\,u_f^2)
\label{collinear-factorisation}
\end{equation}
While the DGLAP approach with fixed order 
coefficient- and splitting functions is 
phenomenologically successful for inclusive quantities like the structure
function 
$F_2(x,Q^2)$ in DIS, it is 
not fully satisfactory from a theoretical point of view, because 
{\it ``the truncation of the
splitting functions at a fixed perturbative order is equivalent to assuming that
the dominant dynamical mechanism leading to scaling violations is the evolution
of parton cascades with strongly ordered transverse 
momenta"} as Catani argued in Ref.~\refcite{catani-feb2000}.
\par
In  the \kt-factorization approach, the partons along the parton ladder are no
longer ordered in transverse momentum. 
At large energies (small $x$) the evolution of parton densities proceeds over a large
region in rapidity $\Delta y \sim \log(1/x)$ and effects of finite transverse
momenta of the partons may become increasingly important.
Cross sections can then be $k_t$ - factorized~\cite{GLR,LRSS2,CCH,CE}
into an off mass-shell ($k_t$ dependent) partonic cross section
$\hat{\sigma}(\frac{x}{z},k_t) $
and a $k_t$ - unintegrated parton density function 
${\cal F}(z,k_t)$:
\begin{equation}
 \sigma  = \int 
\frac{dz}{z} d^2k_t \hat{\sigma}(\frac{x}{z},k_t) {\cal F}(z,k_t)
\label{kt-factorisation}
\end{equation}
The unintegrated gluon density ${\cal F}(z,k_t)$ is 
described by the BFKL
 evolution equation in the region of asymptotically large energies (small $x$). 
\par
 An appropriate description valid for
both, small and large $x$, is given by the CCFM evolution
equation, resulting in an unintegrated gluon density 
${\cal A} (x,\kt,\Pmax ) $, which is a function also of the 
additional evolution scale $\Pmax $ described below. This scale $\Pmax $ is
connected to the factorization scale $\mu_f$ in the collinear approach.
\par
Carrying out the $k_t$ integration in eq.(\ref{kt-factorisation})
explicitly, a form fully consistent with collinear factorization can be 
obtained~\cite{catani-feb2000,catani-dis96}: the
coefficient functions and also the DGLAP splitting functions leading to 
$f_a(z,\mu_f^2)$ are no longer evaluated in fixed order perturbation theory but
supplemented with the all-order resummation of the $\as \log 1/x$ contribution  
at small $x$. 

\section{The CCFM evolution equation}
\label{sec:CCFMEquation}

\begin{figure}[th]
\centerline{\resizebox{0.9\textwidth}{!}{\includegraphics{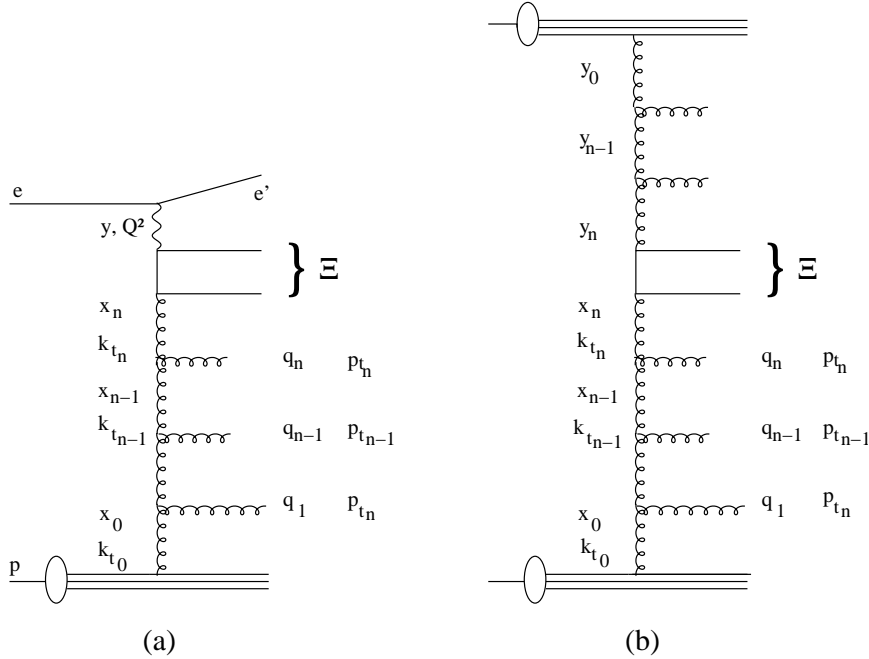}}} 
\vspace*{8pt}
\caption{Kinematic variables for multi-gluon emission in leptoproduction ($a.$)
and hadroproduction ($b$). 
The $t$-channel gluon
four-vectors are given by $k_i$ and the gluons emitted in the initial state
cascade have four-vectors $p_i$. The maximum
angle (a function of the rapidity) for any emission is obtained
from the quark box, as indicated with $\Xi$.
\label{CCFM_variables} } 
\end{figure}
The pattern of QCD initial-state
radiation in a small-$x$ event in $ep$ and $p \bar{p}$ collisions
 is illustrated in
Fig.~\ref{CCFM_variables}  together with labels for the
kinematics.  According to the CCFM evolution equation, the emission of
partons during the initial cascade is only allowed in an
angular-ordered region of phase space. 
 In terms
of Sudakov variables the quark pair momentum is written as:
\begin{equation}
p_q + p_{\bar{q}} = \Upsilon (p^{(1)} + \Xi p^{(2)}) + Q_t
\end{equation}
where $p^{(1)}$ and $p^{(2)}$ are the four-vectors of
 incoming particles (electron-proton or proton-proton), 
respectively and $Q_t$ is the transverse momentum of the quark pair
in the center of mass frame of
$p^{(1)}$ and $p^{(2)}$ (cms).
The variable $\Xi$ is related to the rapidity $Y$ 
in the cms via 
\begin{equation}
Y = \frac{1}{2} \log \left(\frac{E+P_z}{E-P_z}\right)  
= \frac{1}{2}  \log \left(\frac{1}{\Xi}\right)  
\end{equation}
since  $E+p_z=2\Upsilon \sqrt{s}$, $E-p_z=2\Upsilon \Xi\sqrt{s}$ and
$E=\sqrt{s}/2$ with 
$s=(p^{(1)}+p^{(2)})^2$ being the squared center of mass energy.
 Therefore $\Xi$ can be used to define the maximum allowed angle 
in the evolution.
The momenta $p_i$ of the gluons emitted during the initial
state cascade are given by (here treated massless):
\begin{equation}
p_i = \upsilon_i (p^{(1)} + \xi_i p^{(2)}) + \pti{i} \;  , \;\; 
\xi_i=\frac{p_{ti}^2}{s \upsilon_i^2},
\end{equation}
with $\upsilon_i = (1 - z_i) x_{i-1}$ and $x_i = z_i x_{i-1}$. 
The variables 
 $x_i$ and $\upsilon_i$ are the
momentum fractions of the exchanged and emitted gluons, while $z_i$ is
the momentum fraction in the branching $(i-1) \to i$ and $\pti{i}$ is
the transverse momentum of the emitted gluon $i$. 
Again the rapidities $y_i$ are given by $y_i = - 0.5 \log \xi_i $ in the cms. 
\par
The angular-ordered region is then specified by 
(Fig.~\ref{CCFM_variables}$a$ and the lower part of the
cascade in Fig.~\ref{CCFM_variables}$b$, for the upper part  the variables
have to be changed accordingly):
\begin{equation}
\xi_0 < \xi_1< \cdots < \xi_n < \Xi
\end{equation}
which becomes:
\begin{equation}
z_{i-1} q_{i-1} < q_{i} 
\end{equation}
where the rescaled transverse momentum $q_{i}$ of the emitted
gluon is defined by:
\begin{equation}
 q_{i} = x_{i-1}\sqrt{s \xi_i} = \frac{\pti{i}}{1-z_i}
 \label{qbar}
\end{equation}
\par
The CCFM equation for the unintegrated gluon density can be
written~\cite{CCFMd,jung_salam_2000,Salam,Martin_Sutton}
 as an integral equation:
\begin{equation}
{\cal A} (x,\kt,\Pmax ) = {\cal A}_0 (x,\kt,\Pmax ) + \int \frac{dz }{z} 
\int \frac{d^2 q}{\pi q^{2}} \Theta(\Pmax - zq) \Delta_s(\Pmax ,z q) 
\tilde{P}_{gg}(z,q,\kt) {\cal A}\left(\frac{x}{z},\ktp,q\right) 
\label{CCFM_integral} 
\end{equation}  
with $\vec{k}'_{t} = | \vec{k}_{t} + (1-z) \vec{q}|$ and $\Pmax$ being
the upper scale for any emission: 
\begin{equation}
\Pmax > z_n q_n, \; q_n > z_{n-1} q_{n-1},\; \cdots ,\; q_{1} > Q_0
\end{equation}
The
Sudakov form factor $\Delta_s$ is given by:
\begin{equation}
\Delta_s(\Pmax,Q_0) =\exp{\left(
 - \int_{Q_0^2} ^{\Pmax^2}
 \frac{d q^{2}}{q^{2}} 
 \int_0^{1-Q_0/q} dz \frac{\alphasb\left(q(1-z)\right)}{1-z}
  \right)}
  \label{Sudakov}
\end{equation}
with $\alphasb=\frac{3 \as}{\pi}$. For
inclusive quantities at leading-logarithmic order the Sudakov form
factor cancels against the $1/(1-z)$ collinear singularity of the
splitting function.
\par
The splitting function $P_{gg}$ for branching $i$ 
is given by:
\begin{equation}
P_{gg}(z_i,q_i,\kti{i})=  
\frac{\alphasb(\kti{i})}{z_i} \Delta_{ns}(z_i,q_{i},\kti{i}) +
\frac{\alphasb(\pti{i})}{1-z_i}  
\label{Pgg}
\end{equation}
with $\pti{i} = q_i (1-z_i)$ and 
the non-Sudakov form factor $\Delta_{ns}$  defined as:
\begin{equation}
\log\Delta_{ns}(z_i,q_i,\kti{i}) =  -\alphasb
                  \int_{z_i}^1 \frac{dz'}{z'} 
                        \int \frac{d q^2}{q^2} 
              \Theta(\kti{i}-q)\Theta(q-z'q_i)
                  \label{non_sudakov}                   
\end{equation}
The upper limit of the $z'$ integral is constrained by the $\Theta$
functions in eq.(\ref{non_sudakov}) by: $z_i \leq z^{\prime} \leq
\mbox{min}(1,\kti{i}/q_{i}) $, which results in the following form
of the non-Sudakov form factor \cite{Martin_Sutton}:
\begin{equation}
\log\Delta_{ns} = -\alphasb(\kti{i})
\log\left(\frac{z_0}{z_i}\right)
\log\left(\frac{\kti{i}^2}{z_0z_i q_{i}^2}\right)
\label{non_sudakov_int}
\end{equation} 
where
$$z_0 = \left\{ \begin{array}{ll}
    1             & \mbox{if  } \kti{i}/q_{i} > 1 \\
    \kti{i}/\pti{i} & \mbox{if  } z_i < \kti{i}/q_{i} \leq 1 \\
    z_i             & \mbox{if  } \kti{i}/q_{i} \leq z_i  
  \end{array} \right. $$
The non-Sudakov form factor can be written as:
\\
\begin{tabular}{l l l l}
& 
\begin{minipage}[t]{0.09\textwidth}
\vspace*{-0.1cm} 
\rotatebox{0.}{\scalebox{0.38}{\includegraphics*{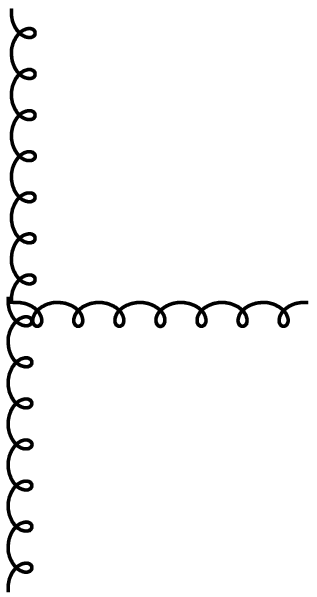}}}
\end{minipage}
& 
\begin{minipage}[t]{0.09\textwidth}
\vspace*{-0.1cm} 
\rotatebox{0.}{\scalebox{0.38}{\includegraphics*{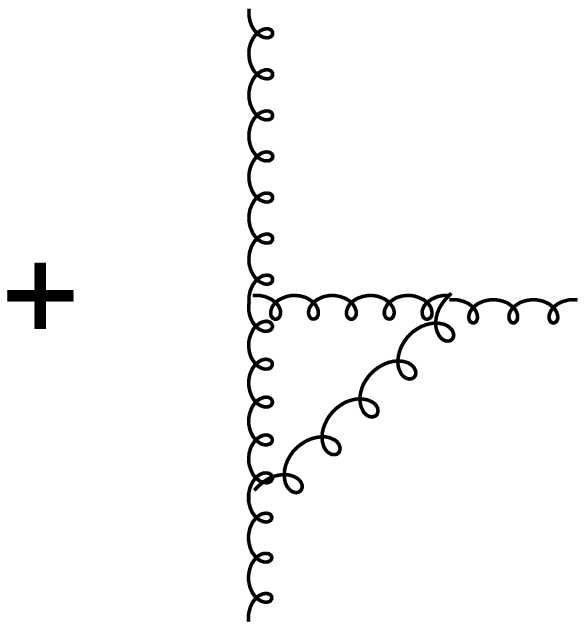}}}
\end{minipage}
& 
\begin{minipage}[t]{0.4\textwidth}
\vspace*{-0.1cm} 
\rotatebox{0.}{\scalebox{0.38}{\includegraphics*{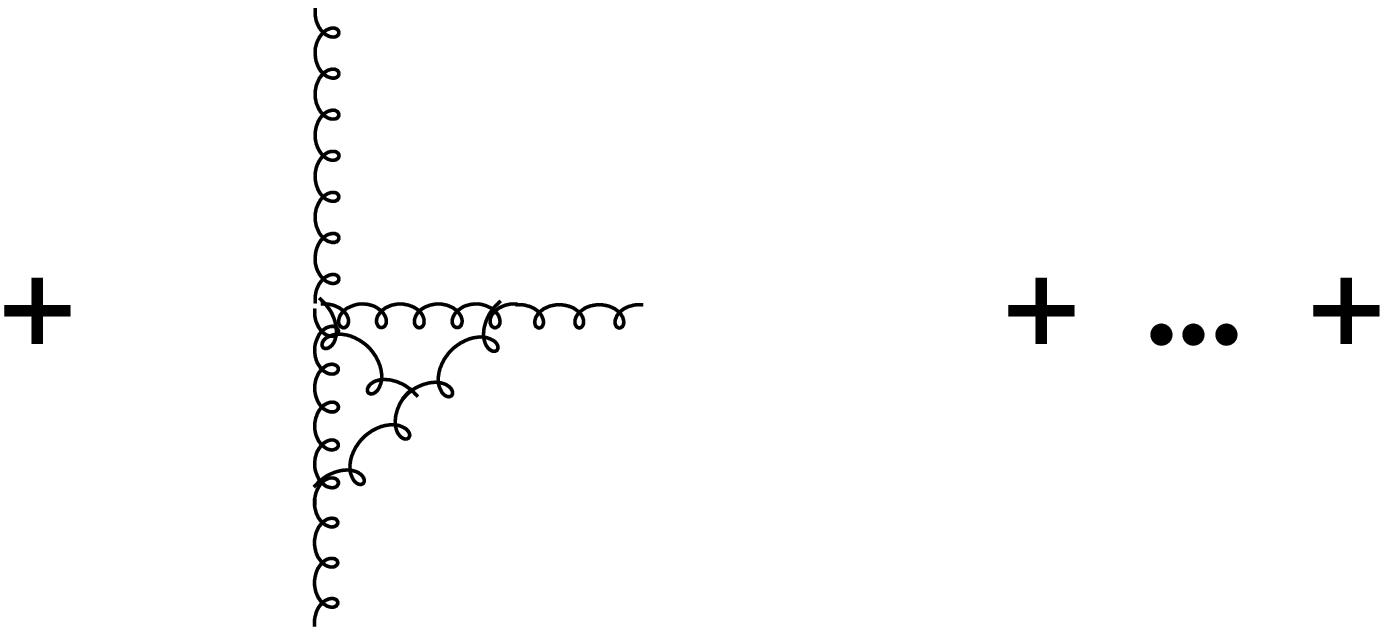}}}
\end{minipage}
\\
& $ \asb \frac{1}{z}$ {\huge $\left[ \right.$} $1 $ & 
$+ \;\;
\asb
\log\left(\frac{z_0}{z_i}\right)
\log\left(\frac{\kti{i}^2}{z_0z_i q_{i}^2}\right)
$ & $ + \;\; \left(\frac{1}{2!}
\asb
\log\left(\frac{z_0}{z_i}\right)
\log\left(\frac{\kti{i}^2}{z_0z_i q_{i}^2}\right)
\right)^2  \;...$
{\huge $\left. \right]$} 
\end{tabular}
\vspace*{0.8cm}\\
where the similarity with the Sudakov form factor becomes obvious. Note however,
that the Sudakov form factor $\Delta_s$ resums the large $z$ contributions,
whereas the non-Sudakov form factor $\Delta_{ns}$ resums  the
small $z$ ones. 
\par 
In the CCFM approach 
the scale $\Pmax$ (coming from the maximum angle) 
can be written as (see eq.(\ref{qbar})):
\begin{equation}
 \Pmax^2 = \Upsilon^2 \Xi s 
 = \hat{s} + Q_{\prp}^2 
\end{equation}
with $\hat{s}=(p_q + p_{\bar{q}})^2$ and the relation of $\Pmax$
to a particular choice of the factorization scale 
$\mu_f$ in the collinear approach becomes obvious.
\subsection{Improvements to the CCFM splitting function} 
Originally, 
the CCFM splitting function $P_{gg}$, as given in eq.(\ref{Pgg}), included 
only the soft and collinear singular terms. In the
asymptotic region of large energies this is a reasonable choice, but effects
from non-leading contributions are expected at presently accessible energies. 
For simplicity also
the scale in \as was chosen differently in the small and large
$z$ part of the splitting function (see eq.(\ref{Pgg})).
In
Ref.~\refcite{smallx_2001} it was suggested to use $\mu_r = \pt=\qb (1-z)$ 
everywhere and
 to include the non-singular terms in the splitting function. These changes
are non-trivial as they need to be considered both in the Sudakov and
non-Sudakov form factors.
\par
In the following improvements to the
originally proposed CCFM splitting function are discussed.
\subsubsection{The soft region}
In a DGLAP type evolution  with the transverse momenta of the gluon propagators
increasing from the proton towards the hard scattering, the non-perturbative
region with  $\kt < \kt^{cut}$  has an influence on the initial parton
density only. In CCFM, due to angular ordering 
a kind of random walk in the propagator gluon 
\kt\  can be performed. Even during the evolution the non-perturbative region
can be entered for $\kt < \kt^{cut}$. 
In the region of small \kt ,
 $\as$ and the parton density are large, 
and collective phenomena, like gluon recombination or saturation might play a
role. Thus, the fast increase of the parton density and the cross section is
tamed. 
Much effort was put recently 
into a more detailed understanding of this special region of
phase space (for example see 
Ref.~\cite{Kovchegov:99,Balitsky:96,smallx_2002,GBS:03}). 
However, for the calculation of the unintegrated gluon density
presented here, a simplified but
practical approach is taken:   $\as(\mu)$ is fixed 
for $\mu < Q_0 $ and  no emissions are allowed  
until $\kt > \kt^{cut}$ is reached.
\subsubsection{The non-singular terms in $P_{gg}$}
The  $P_{gg}$ splitting function  used in the collinear approach  contains also
non-singular terms.
Such non-singular terms can be included in the CCFM splitting function, but
care has to be taken, which terms $\Delta_{ns}$ is acting 
on to ensure positivity of $P_{gg}$~\cite{smallx_2001} :
\begin{equation}
P_{gg} = \asb \left(\kt \right) 
\left( \frac{(1-z)}{z} + \frac{z(1-z)}{2}\right) \Delta_{ns} 
  + \asb (\pt) \left(\frac{z}{1-z} + 
\frac{z(1-z)}{2}\right) 
\label{Pgg_full}
\end{equation}
As the splitting function is also part of $\Delta_{ns}$ and $\Delta_{s}$,
they need to be modified accordingly~\cite{smallx_2001}. The non-Sudakov
form factor including the full splitting function is then given by:
\begin{equation}
\log\Delta_{ns} =  -\asb \left(\kt \right)
                  \int_0^1 dz
                        \left( \frac{1-z}{z} + \frac{z(1-z)}{2} \right)
                        \int \frac{d {q}^2}{{q}^2} 
              \Theta(\kt-q)\Theta(q-zq)
\label{non_sudakov_full}
\end{equation}

\subsubsection{The scale in $\as$} 
It was suggested in Ref.~\refcite{smallx_2001}, 
to change the scale in $\alpha_s$  to $\pt=\qb (1-z)$ 
also in the small $z$ part of the
splitting function ${P}_{gg}$:
\begin{equation}
{P}_{gg}=  
\frac{\alphasb(\pt)}{z} \Delta_{ns} + \frac{\alphasb(\pt)}{1-z}
\label{Pgg_asq}
\end{equation}
 As a consequence, the non-Sudakov form factor,  
changes from eq.(\ref{non_sudakov_int}) 
to~\cite{smallx_2001,jung-dis02}:
\begin{equation}
\log\Delta_{ns}                = -\int_0^{1} \frac{dz}{z}
                  \int^{k_{t}^2}_{(z\qb)^2)} 
			\frac{d q^2}{q^2}
                  \frac{1}{\log(q/\Lambda_{qcd})}
		\label{non_sudakov_asq} 	
\end{equation}
It is obvious, that a special treatment of the soft region is needed,
because $q$ can become very small, even 
$q < \Lambda_{qcd}$  at small values of $z$. 
However, in practical applications we observe only a
small effect when changing the scale of the small $z$ part from \kt\ to \pt.
\subsection{The unintegrated gluon density}\label{ccfmfits}
\begin{figure}[th]
\vskip -0.5cm
\centerline{\resizebox{0.9\textwidth}{!}{\includegraphics{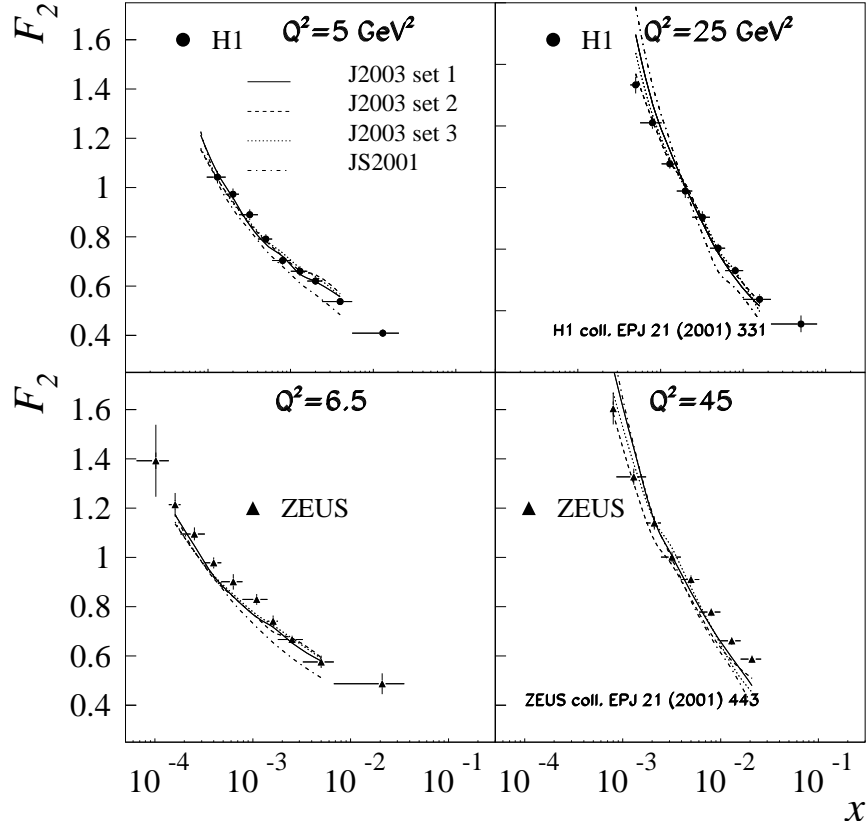}}} 
\caption{The structure function $F_2(x,Q^2)$ as measured by H1~\protect\cite{H1_F2_2001}
and ZEUS~\protect\cite{ZEUS_F2_2001} together with the results of the different fits
described in Tab.~\protect\ref{pdfsets}
\label{f2-fits} } 
\end{figure}

The CCFM evolution equations have been solved 
numerically~\cite{jung_salam_2000} using a Monte Carlo
method\footnote{A Fortran program for the unintegrated gluon
density  $x {\cal A}(x,k_{t},\Pmax)$ can be obtained from 
Ref.\refcite{CASCADEMC}.}. 
\begin{figure}[htb]
\vskip -0.5cm
\centerline{\resizebox{0.9\textwidth}{!}{\includegraphics{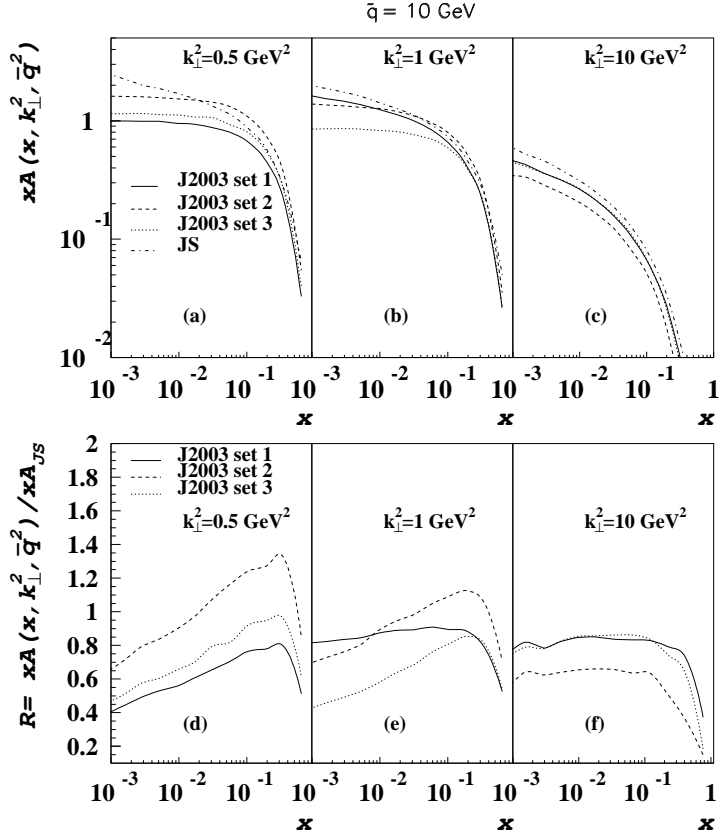}}} 
\vskip -0.5cm
 \caption[*]{
 {\it Comparison of the different sets of unintegrated gluon densities obtained
 from the CCFM evolution as described in the text. In $(a-c)$ the unintegrated
 gluon density is shown as a function of $x$ for different values of \kt\ at a
 scale of $\bar{q}=10$ GeV. In $(d-f)$ the ratio 
 $R=\frac{x{\cal A}(x,\kt^2,\Pmax^2)}
 {x{\cal A}(x,\kt^2,\Pmax^2)_{\it JS}}$ as a function of $x$ 
 for different values of \kt\  is shown.
 \label{ccfm-new}}}
\end{figure}
The unintegrated gluon density at any $x$, $\kt$ and scale $\Pmax$ is obtained
by evolving 
a starting gluon distribution from the scale $Q_0$ according to CCFM to the
scale $\Pmax$. 
The normalization $N$ of
the input distribution as well as the starting scale $Q_0$, which also acts as a
collinear cutoff to define $z_{max} = 1- Q_0/q$, need to be specified. 
These parameters were  
fitted such that 
the structure function $F_2$ as measured at 
H1~\cite{H1_F2_1996,H1_F2_2001} and ZEUS~\cite{ZEUS_F2_1996,ZEUS_F2_2001}
can be described after convolution with the off-shell matrix element
in the region of $x < 5\cdot 10^{-3}$ and $Q^2 > 4.5$~GeV$^2$. 
According to the discussion in the last section, the following sets of  
CCFM unintegrated gluon densities are obtained: 
\begin{itemlist}
\item[$\bullet$] {\it JS2001} (Jung, Salam~[\refcite{jung_salam_2000}])\\
  The splitting function $P_{gg}$ of eq.(\ref{Pgg}) is used, with $Q_0=1.4$~GeV.
  The soft region is defined by $\kt^{cut} = 0.25 $ GeV.
\item[$\bullet$] {\it J2003 set 1} [\refcite{jung-dis03}]\\
  The CCFM splitting function containing only singular terms 
  (eq.(\ref{Pgg})) 
   is used, with $\kt^{cut}=Q_0$ fitted to $\kt^{cut} = Q_0=1.33$~GeV. 
\item[$\bullet$] {\it J2003 set 2} [\refcite{jung-dis03}] \\
  The CCFM splitting function (eq.(\ref{Pgg_full})) 
  containing also the non singular terms is used. 
  The Sudakov and non-Sudakov form
  factors were changed accordingly. 
  The collinear cut is fitted to $Q_0=\kt^{cut} = 1.18$ GeV.
\item[$\bullet$] {\it J2003 set 3} [\refcite{jung-dis03}] \\
  CCFM splitting function containing only singular terms 
   but the scale in $\alpha_s$ is changed from \kt\ to \pt\ for the
  $1/z$ term.  The collinear cut is fitted to $Q_0=\kt^{cut} =1.35$ GeV.
  The problematic region in the non-Sudakov form factor in 
  eq.(\ref{non_sudakov_int}) is avoided by fixing $\alpha_s(\mu)$ for 
  $\mu<0.9$ GeV.
\end{itemlist}
A comparison of the different sets of CCFM unintegrated gluon densities is shown
in Fig.~\ref{ccfm-new}. It is clearly seen, that the treatment of the soft
region, defined by $\kt < \kt^{cut}$ influences the behavior at small $x$ and
small \kt. After convolution with the off-shell matrix
elements, all sets describe the structure function $F_2(x,Q^2)$ 
reasonably well. 
In Fig.~\ref{f2-fits} results of the fits are compared with measurements of the
structure function $F_2(x,Q^2)$ as obtained by the H1~\cite{H1_F2_2001} and 
ZEUS~\cite{ZEUS_F2_2001} collaborations.
In Tab.~\ref{pdfsets} the parameters of the CCFM unintegrated gluon densities
are summarized indicating  also the quality of the fits. It is interesting to
note, that the quality of the CCFM fits  is similar to that 
obtained from   
global fits in the collinear DGLAP approach (e.g. Ref.~\refcite{MRST,CTEQ6}).
\begin{table}[h]
\tbl{Parameters of the CCFM unintegrated  gluon densities.
The $\chi^2/ndf$  is obtained from a comparison to HERA 
$F_2$ data~\protect\cite{H1_F2_1996,H1_F2_2001,ZEUS_F2_1996,ZEUS_F2_2001}
in the range  $x< 5\cdot 10^{-3}$ and $Q^2 > 4.5 $ GeV$^2$. 
\label{pdfsets}}
{
\begin{tabular}{@{}cccccc@{}}
\toprule
set & $P_{gg}$ & $\Delta_{ns}$ & $Q_0$ (GeV)  & $\kt^{cut}$ (GeV) & $\chi^2/ndf$  \\
\colrule
{\it JS2001}~\protect\cite{jung_salam_2000}          
                  &  eq.(\ref{Pgg})& eq.(\ref{non_sudakov_int})  
			& 1.40  &  0.25  & 1197/248 = 4.8\\
{\it J2003 set 1} &  eq.(\ref{Pgg})&eq.(\ref{non_sudakov_int})  
                  & 1.33 & 1.33 & 321/248 = 1.29\\
{\it J2003 set 2} &  eq.(\ref{Pgg_full}) & eq.(\ref{non_sudakov_full})
                  & 1.18  &  1.18  & 293/248 = 1.18\\
{\it J2003 set 3} & eq.(\ref{Pgg_asq}) &eq.(\ref{non_sudakov_asq})  
                  & 1.35  & 1.35  & 455/248 = 1.83 \\ 
\botrule
\end{tabular}}
\end{table}

\section{Comparison with hadronic final state data}
A comparison of measurements of hadronic final
state properties, like jet or heavy quark cross sections, with theoretical
predictions requires a detailed simulation of the experimentally accessible
phase space. Such simulations are provided by Monte Carlo event generators,
which also allow to apply the hadronization step. Monte Carlo event generators
for DGLAP type collinear factorized processes are widely used (e.g. 
\pythia \cite{\PYTHIAMC}, \rapgap \cite{\RAPGAPMC}, 
\herwig \cite{\HERWIGMC}). Two Monte Carlo generators 
 have been developed
including the small $x$ evolution equations: 
\cascade \cite{\CASCADEMC}, which 
follows explicitely the CCFM approach of angular ordering and 
\ldcmc \cite{\LDCMC} which is a reformulation of CCFM (not described here). 
It has been
shown \cite{jung_salam_2000}, that the parton shower approach used in \cascade\  
reproduces exactly the properties of the CCFM evolution described in 
section~\ref{sec:CCFMEquation}. 
The \cascade~Monte Carlo event generator 
has been frequently used for comparison with HERA
measurements, like heavy quark ~\cite{\hqhera} and  
high \pt ~jet~\cite{\jetshera} production,
and also with bottom
production at the Tevatron~\cite{jung-hq-2001}.
\par
In the following sections a few examples  
are presented where data are compared
to predictions obtained with \cascade\ 
based on the new unintegrated gluon
densities, described in section ~\ref{ccfmfits}.
\subsection{Jet cross section at HERA}
\begin{figure}[ht]
\begin{minipage}{0.45\textwidth}
\vskip -1.0cm
\centerline{\resizebox{1\textwidth}{!}{\includegraphics{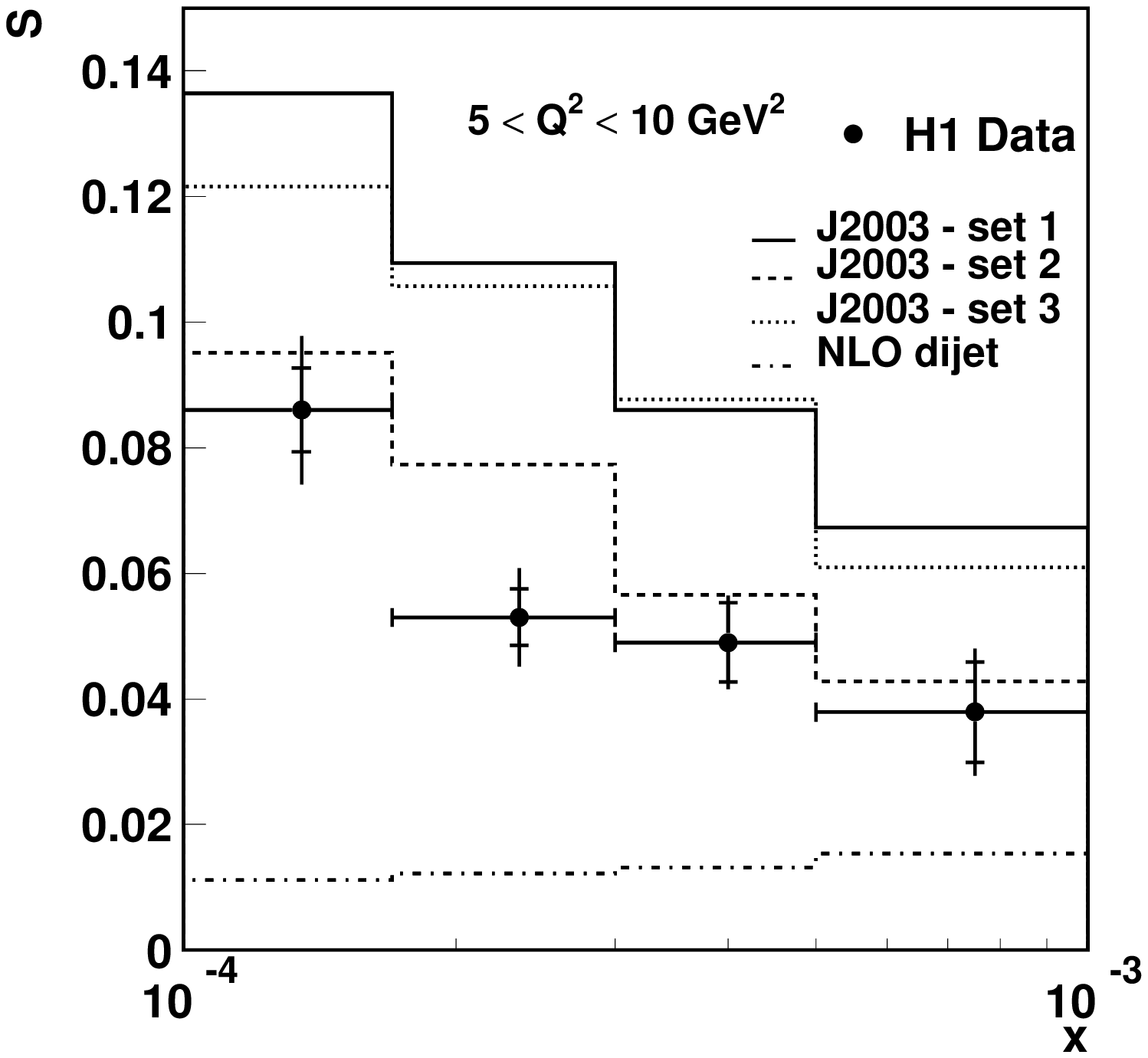}}} 
\vskip -0.2cm
\caption{{ The $S$ distribution for dijet events\protect\cite{H1dijet}. 
 The prediction from 
\cascade~  using {\it J2003 set 1} and {\it J2003 set 2} are shown together
with the NLO-dijet prediction in the collinear factorization approach.
   }\label{azim:casc-nlo}}
\end{minipage}
\hspace*{0.3cm}
\begin{minipage}{0.45\textwidth}
\vskip -1.0cm
\centerline{\resizebox{1.1\textwidth}{!}{\includegraphics{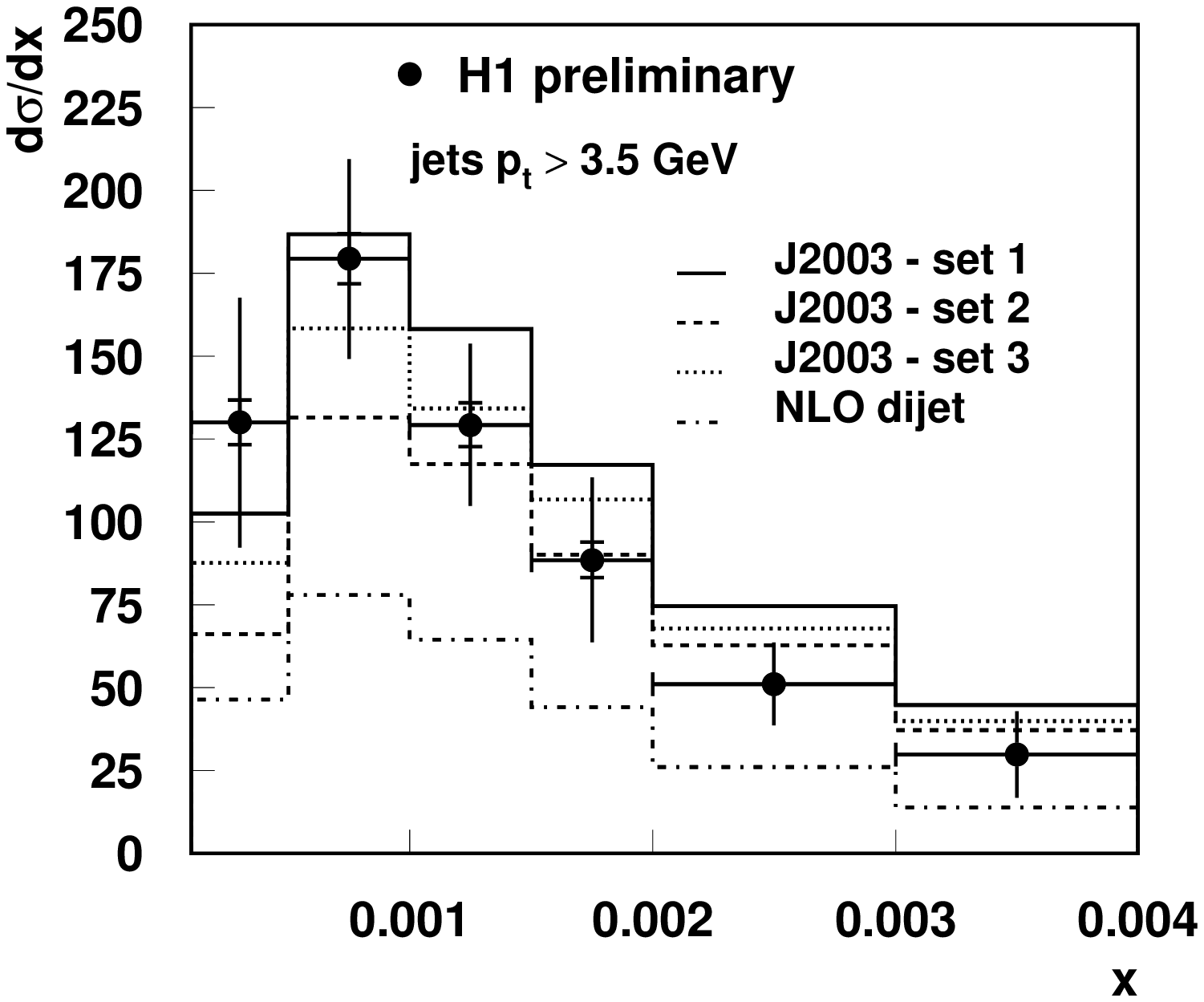}}} 
\vskip -0.5cm
\caption{The cross section for forward jet production as measured by 
H1\protect\cite{H1fwdjet2} as a function of \xbj. The prediction are the same as
in Fig.~\protect\ref{azim:casc-nlo}.
\label{fjet:cas-nlo} } 
\end{minipage}   
\end{figure}

The azimuthal correlation of dijets at HERA is sensitive
to the transverse momentum of the partons incoming to the hard scattering
process and therefore sensitive to the details of the unintegrated gluon 
density. This was studied in a measurement~\cite{H1dijet} of 
the cross section for dijet production with $E_T > 5 (7)$ GeV in the
range $1 < \eta_{lab} < 0.5$ in deep-inelastic scattering 
 ($10^{-4} < x < 10^{-2}$, $ 5 < Q^2 < 100$ GeV$^2$). 
In LO collinear factorization,  dijets at small
$\xbj$ are produced essentially by 
$\gamma g \to q\bar{q}$, with the gluon collinear to the
incoming proton. Therefore the $q\bar{q}$ pair is produced back-to-back in the
plane transverse to the $\gamma^* p$ direction. 
From NLO (${\cal O}(\alpha_s^2)$) on,  significant deviations
from the back-to-back scenario can be expected. In the \kt-factorization 
approach the transverse momentum
of the incoming gluon, described by the unintegrated gluon density,
is taken explicitely into account, resulting in deviations from a
pure  back-to-back configuration. 
The azimuthal de-correlation, as suggested in Ref.~\refcite{Szczurek}, can be
measured:
\begin{equation}
 S = \frac{\int_{0}^{\alpha}N_{2-jet}(\Delta\phi^{*},x,Q^2)d\Delta\phi^{*}}
            {\int_{0}^{180^{\dg}}N_{2-jet}(\Delta\phi^{*},x,Q^2)d\Delta\phi^{*}}, 
                0 < \alpha < 180^{\dg} 
\end{equation}
In the measurement shown in Fig.~\ref{azim:casc-nlo},
$\alpha = 120^{\dg}$ has been chosen. The data are compared to 
predictions from \cascade\ using {\it J2003 set 1 - 3}. 
Also shown for comparison is the NLO-dijet~\cite{DISENT} calculation of the
collinear approach. 
One clearly sees, that a fixed order NLO-dijet calculation is not sufficient, 
whereas {\it J2003 set 2} gives a good description of the data.
However, the variable $S$ is  sensitive to  the details of the unintegrated
gluon distribution, as can be seen from the comparison with 
{\it J2003 set 1} and {\it set 3}.
\par
A measurement, aiming to observe deviations from the
collinear DGLAP approach, is the production of jets in the forward (proton)
region. The phase space is restricted to a region of $Q^2 > 5$ \gevsq~and 
$E_T^{jet} > 3.5$ \gev~in the forward region of 
$1.7 <\eta_{jet} < 2.8$ 
with the additional requirement of 
$0.5 < E^2_{{T,jet}}/Q^2 < 2$, a region where the 
 contribution from the evolution in $Q^2$ is small. 
The cross section for forward jet production has been measured by 
H1~\cite{H1fwdjet2} as a function of \xbj , shown in
Fig.~\ref{fjet:cas-nlo} together with predictions from \cascade\ using 
{\it J2003 set 1 - 3}. Also the NLO-dijet~\cite{DISENT}
prediction 
in the collinear approach is shown.
The fixed NLO-dijet calculation falls below the measurement, whereas the
\kt-factorization approach supplemented with CCFM evolution gives a reasonable
description of the data for all {\it J2003 set 1 - 3}.

\subsection{Charm Production at the Tevatron}
\begin{figure}[ht]
\vskip -0.5cm
\centerline{\resizebox{0.9\textwidth}{!}{\includegraphics{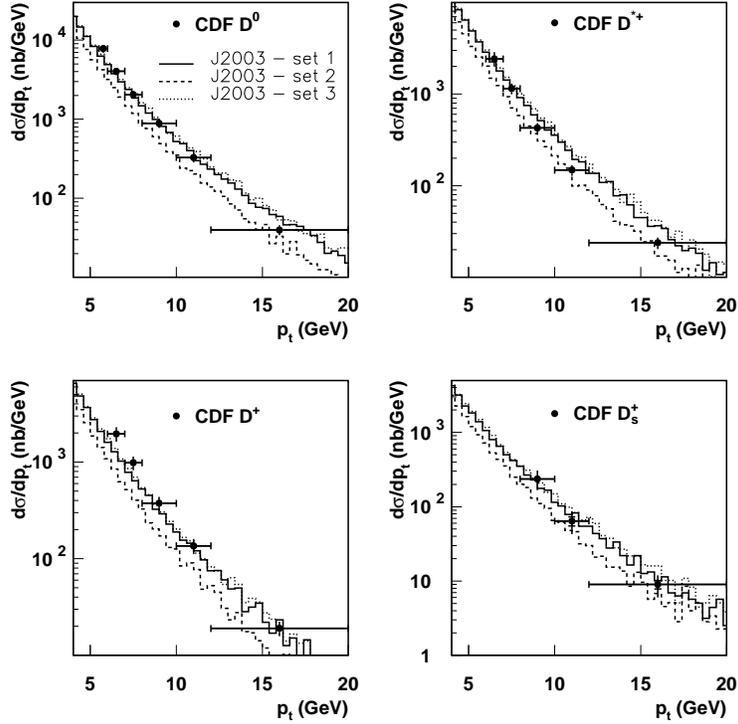}}} 
\vskip -0.5cm
\caption{ Differential cross section for $D$ meson production as measured by 
CDF~\protect\cite{CDFcharm} as a function of the transverse momentum compared to
predictions from \cascade .
\label{charmcdf}} 
\end{figure}

The differential cross section as a function of the transverse momentum of
$D$-mesons  
has been measured in $p\bar{p}$ collisions 
at $\sqrt{s}=1.96$~TeV by the
CDF collaboration~\cite{CDFcharm}. They find the measured 
cross section to be larger than the NLO 
predictions in the collinear factorization approach by about 100 \% at low \pt\ 
and 50 \% at high \pt. 
In Fig.~\ref{charmcdf} the measurement is shown together with
the predictions obtained in \kt-factorization using 
\cascade\ with the CCFM unintegrated gluon density described above. 
For {\it J2003 set 1} and {\it set 3} good agreement for all 
measured charmed mesons is observed. The cross section predicted 
using {\it J2003 set 2} falls below the measurement, in
contrast to the observation in Fig.~\ref{azim:casc-nlo}, showing  
the sensitivity of the measurements on the details of the unintegrated gluon
density.
\subsection{Bottom Production at the Tevatron}
The cross section for $b \bar{b}$ production 
in $p\bar{p}$ collision at
$\sqrt{s}=1800$~GeV has also been compared
with the prediction of \cascade\ based on the CCFM gluon densities.
\begin{figure}[htb]
  \vspace*{2mm}
  \begin{center}
\begin{minipage}{0.45\textwidth}
  \begin{center}
\centerline{\resizebox{1\textwidth}{!}{\includegraphics{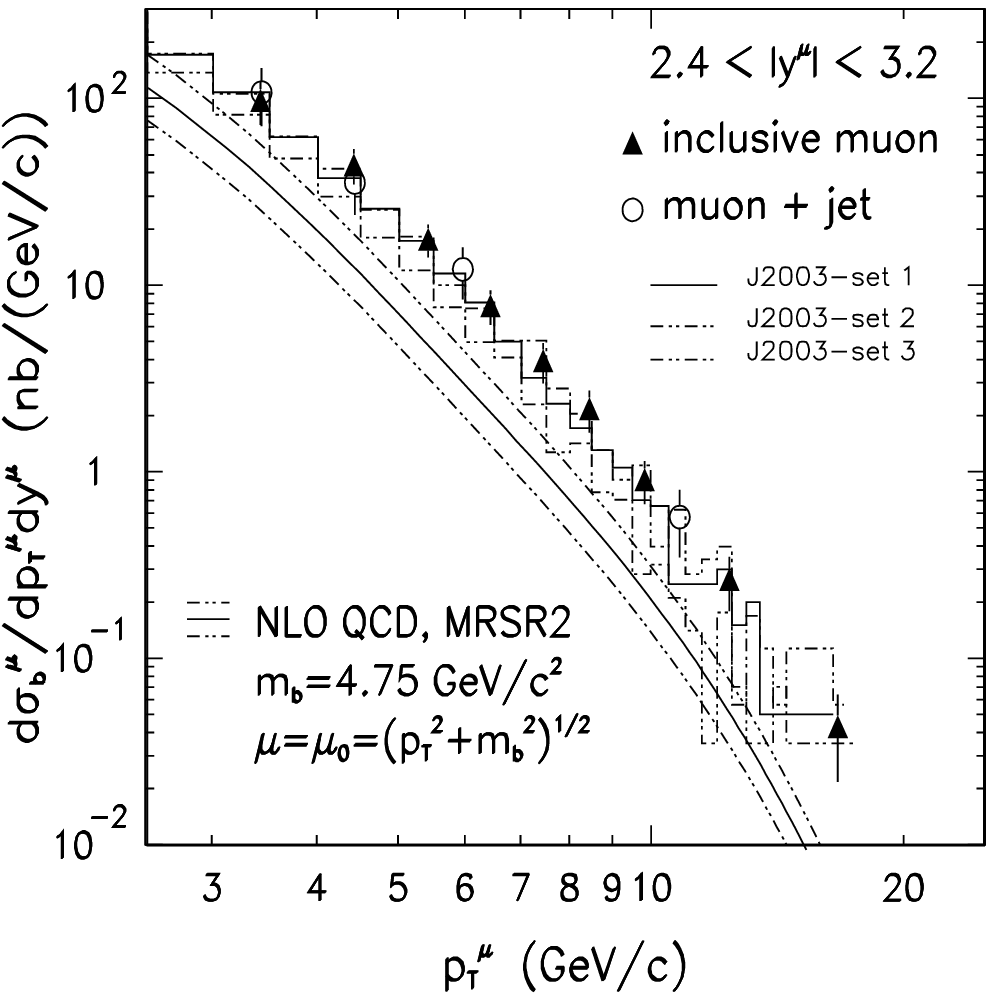}}} 
\vskip -0.2cm
\caption{{  
Cross section for muons from $b$-quark decays as a function of $p_T^{\mu}$ (per
unit rapidity) as measured by  D0~\protect\cite{d0_bb_muon} compared to the
prediction of \cascade .
    }\label{d0_ptmuon}}
    \end{center}
\end{minipage} 
\hspace*{0.5cm}
\begin{minipage}{0.45\textwidth}
\centerline{\resizebox{1\textwidth}{!}{\includegraphics{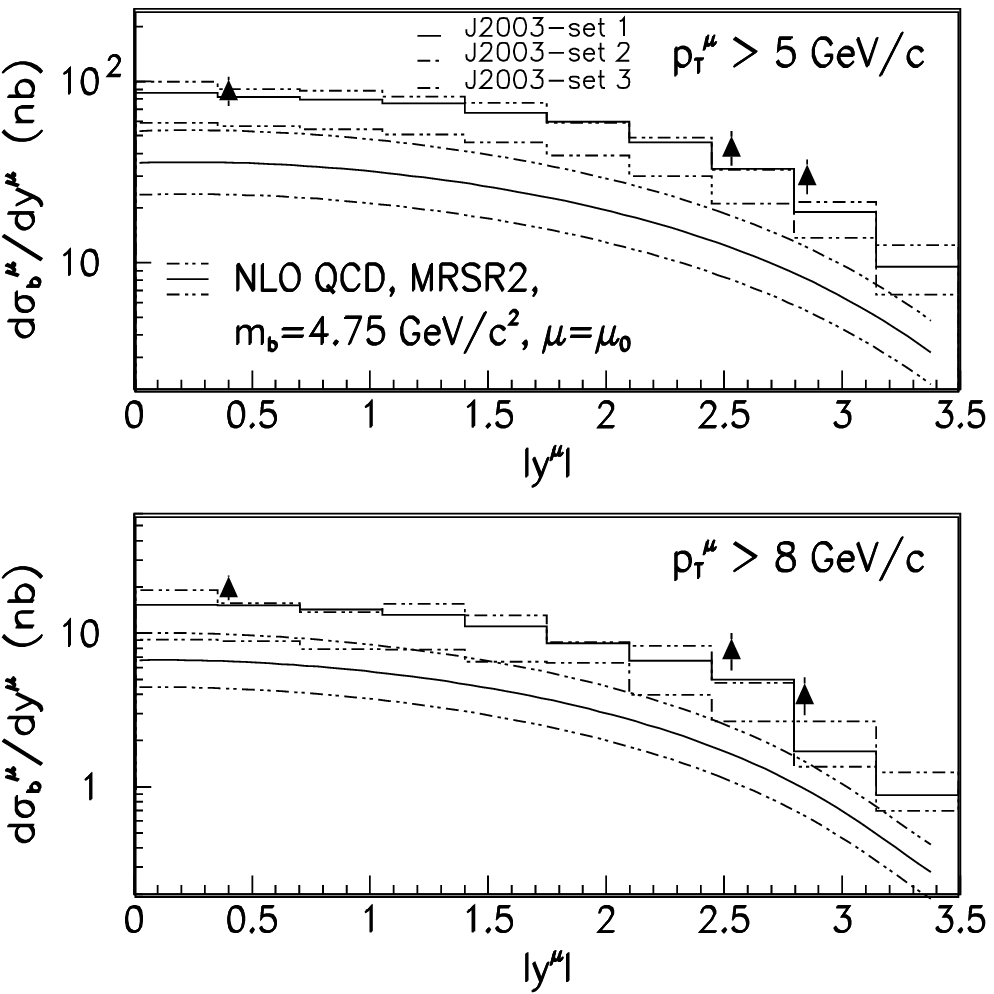}}} 
\vskip -0.2cm
\caption{{  
Cross section for muons from $b$-quark decays as a function of $|y^{\mu}|$ 
for two different $p_T^{\mu}$ cuts 
as measured by  D0~\protect\cite{d0_bb_muon} compared to the
prediction of \cascade .
    }\label{d0_etamuon}}
\end{minipage}   
    \end{center}
\end{figure}
Since \cascade~ generates full hadron level events, direct comparisons with
measured cross sections 
for the production of $b$ quarks decaying semi-leptonically into muons
are possible.
The muon cross sections as a function of the transverse momentum $p_T^{\mu}$ 
and pseudo-rapidity $|y^{\mu}|$ 
as measured by D0~\cite{d0_bb_muon} are compared to the \cascade~ prediction in  
Fig.~\ref{d0_ptmuon} and Fig.~\ref{d0_etamuon}. Both the $p_T^{\mu}$ and 
$|y^{\mu}|$ cross sections are well described with {\it J2003 set 1}
and {\it set 3}.
The prediction based on {\it J2003 set 2} falls below the measurement, as was
already observed in the charm case, again indicating the sensitivity to the
details of the unintegrated gluon density.

\subsection{Higgs production in $p\bar{p}$}
\begin{figure}[ht]
\begin{minipage}{0.95\textwidth}
\vskip -1.cm
\centerline{\resizebox{0.95\textwidth}{!}{\includegraphics{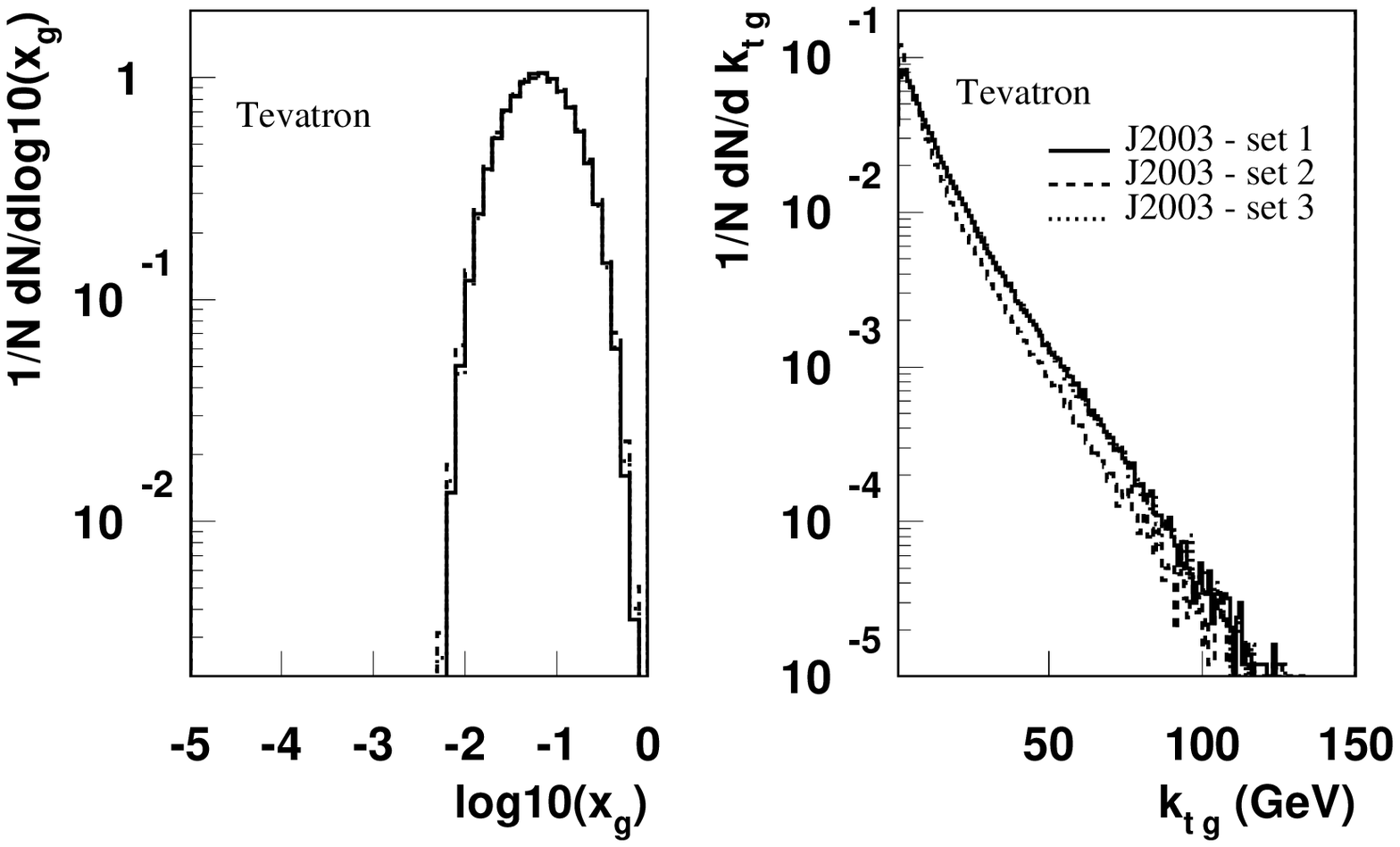}}} 
\vskip -2.cm
\centerline{\resizebox{0.95\textwidth}{!}{\includegraphics{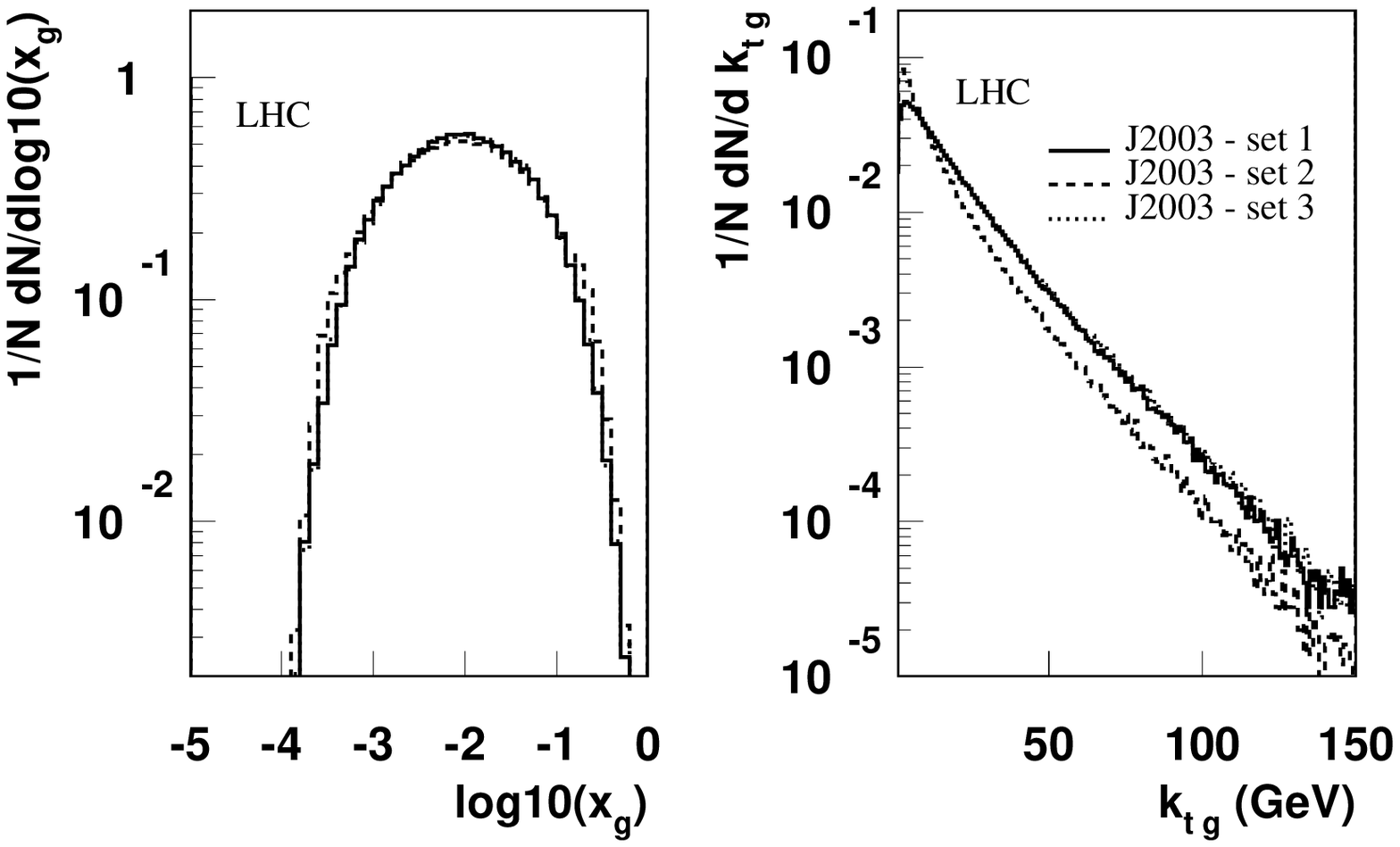}}} 
\vskip -1.cm
\caption{ Longitudinal and transverse momenta of the gluons for $ g^* g^* \to h$
for Tevatron (upper) and LHC (lower) energies.
\label{x-pt-higgs}} 
\end{minipage}
\end{figure}

\begin{figure}[ht]
\begin{tabular}{c c}
\begin{minipage}{0.50\textwidth}
\vskip -0.5cm
\centerline{\resizebox{0.95\textwidth}{!}{\includegraphics{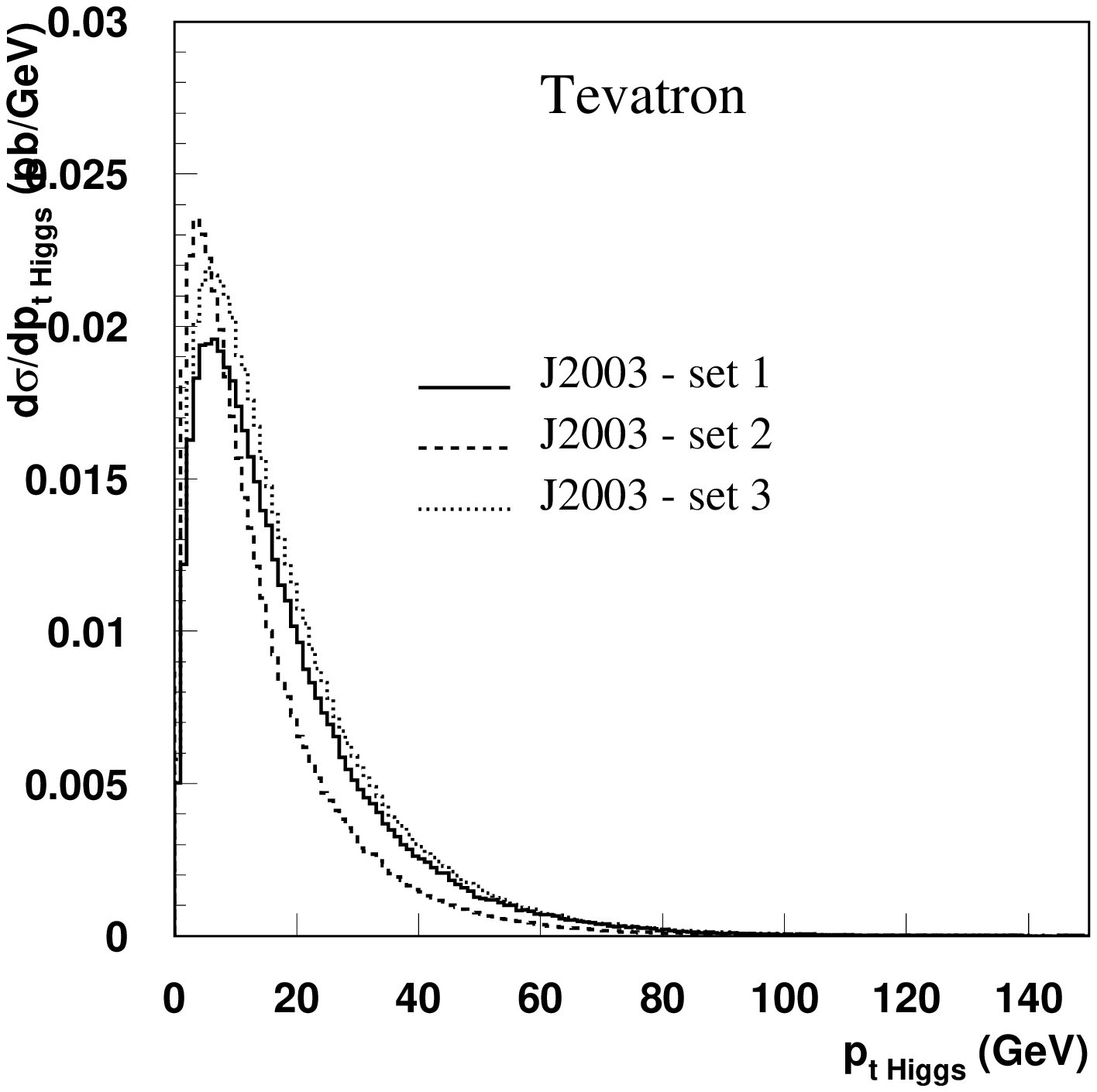}}} 
\vskip -0.5cm
\end{minipage}&
\begin{minipage}{0.50\textwidth}
\vskip -0.5cm
\centerline{\resizebox{0.95\textwidth}{!}{\includegraphics{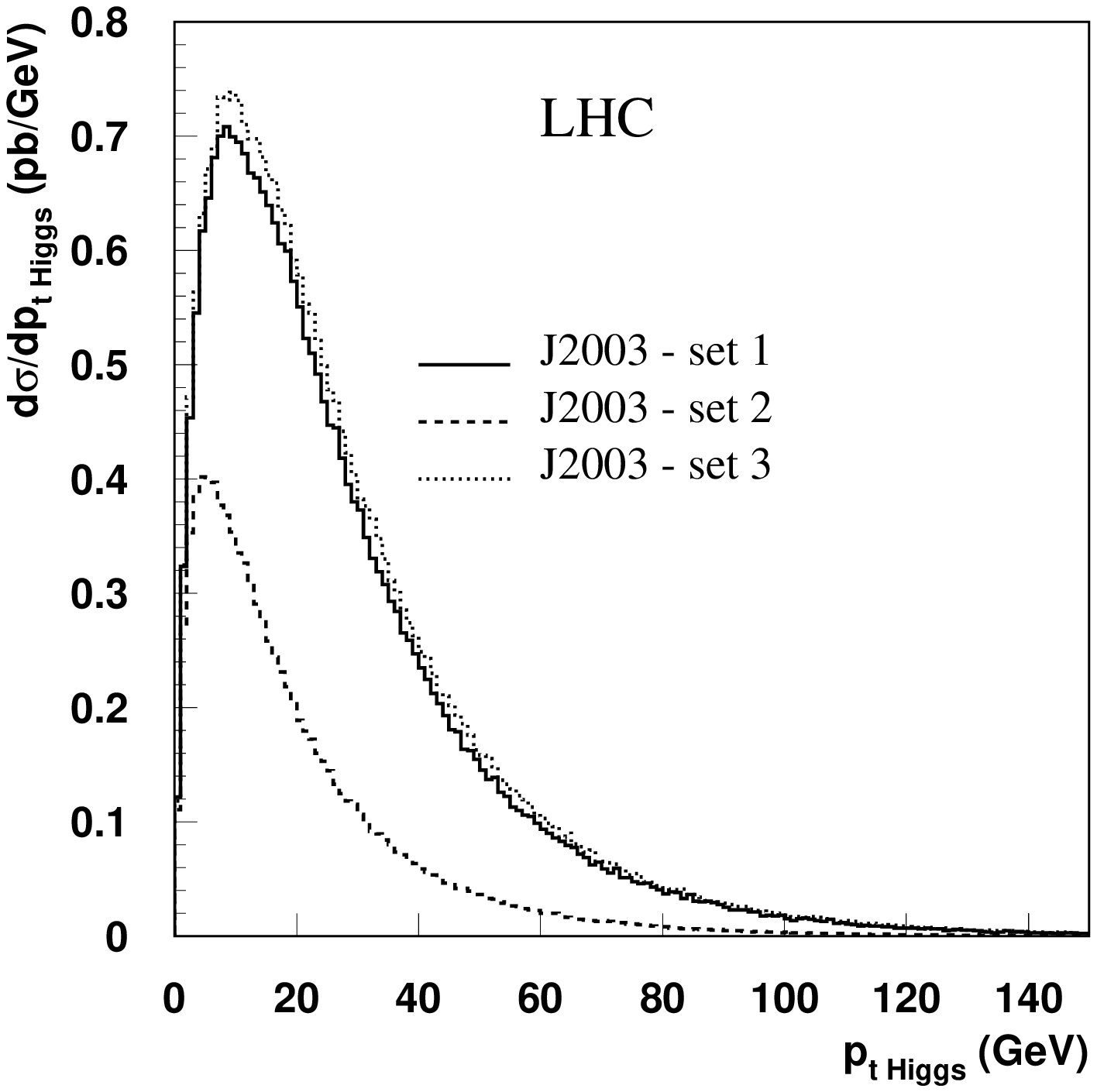}}} 
\vskip -0.5cm
\end{minipage}
\end{tabular} 
\caption{ Differential cross section for 
Higgs production as a function of the transverse momentum 
for $m_{Higgs} = 125$ GeV obtained with the different un-integrated gluon
densities.
\label{higgs-pt}} 
\end{figure}

Higgs production at high energies proceeds predominantly 
via gluon - gluon scattering. Much effort has been put into the calculation of
higher order corrections in the collinear approach, not only for 
the calculation of the
total production cross section, but also for 
the calculation of the transverse momentum spectrum of
the Higgs boson~\cite{catani-qt,stirling-higgs,catani-nll}. 
At the large Tevatron or LHC energies
 and with an expected 
Higgs mass of ${\cal O}(100 - 200) $ GeV, the \kt-factorization approach
can be also used to estimate higher order corrections. 
The off-mass-shell matrix element $g^* g^* \to h$
has been calculated in Ref.~\refcite{hautmann-higgs} in the high energy
approximation  for $m_t \to \infty$ and is now implemented in \cascade .
In Fig.~\ref{x-pt-higgs} the distribution of the longitudinal momentum 
fractions $x$ of the gluons and their 
transverse momenta are 
shown for Tevatron and LHC energies. In both cases, the longitudinal momenta
reach values of the same order as the average
transverse momenta ($\sim 10 (18)$ GeV for
Tevatron (LHC) energies, respectively), 
making the \kt-factorization approach applicable.
It can be seen in Fig.~\ref{x-pt-higgs}, that the \kt\ spectrum of the gluons is
different for the different unintegrated gluon densities
described in sec.~\ref{ccfmfits}, both in the small and also large \kt region.
In Fig.~\ref{higgs-pt} the differential cross section for Higgs production
as a function of $p_t$ obtained with \cascade\ is shown for 
the different unintegrated gluon densities. 
The total cross section for Higgs production 
is given in Tab.~\ref{xsect-higgs}.
It is interesting to note, 
that the different set of unintegrated gluon densities
predict similar cross sections as a function of \pt\   
for Tevatron energies whereas at LHC the cross
section differs by factors up to 3. This again clearly indicates the sensitivity
to the details of the unintegrated gluon density, which can be determined much
more precisely with the forthcoming measurements at HERA. 
A similar
result has been obtained in Ref.~\refcite{kwiecinski-higgs,kwiecinski-ccfm}, 
approximating   
the matrix element and the CCFM unintegrated gluon distribution and also
applying doubly unintegrated parton distributions in Ref.~\refcite{Watt:2003vf}.
\begin{table}
\begin{minipage}{1\textwidth}
\tbl{Cross section for Higgs boson production $\sigma (pp \to h X) $
\label{xsect-higgs}}
{%
\begin{tabular}{@{}c    c    c @{}}
\toprule
 gluon distribution                 &  \multicolumn{2}{c}{$\sqrt{s}$ }   \\
                        &  1.96 TeV      &  14.0  TeV  \\
 \colrule 
 {\it J2003 set 1}      &   0.47 pb       &   22.6  pb    \\
 {\it J2003 set 2}      &   0.38 pb       &    7.8 pb      \\
 {\it J2003 set 3}      &   0.54  pb    &     24.6 pb    \\
 \botrule
\end{tabular}}
\end{minipage}
\end{table}
\section{Conclusion}
It has been shown, that \kt-factorization and the CCFM evolution of the gluon
density is a powerful tool for the description of hadronic final state
measurements. The unintegrated gluon densities, 
which are obtained from CCFM evolution convoluted with the off-mass-shell 
matrix elements to describe HERA $F_2$ data
are implemented in the full hadron level Monte Carlo event generator
\cascade .
\par
Jet measurements at HERA, but also measurements of charm and bottom production
at the Tevatron can be reasonably well described (with specific sets of
unintegrated gluon densities),  
whereas calculations performed
in the collinear approach even in NLO have difficulties to describe the data. 
This shows the
advantage of applying \kt-factorization to estimate higher order contributions
to the cross section  but also the
importance of a detailed understanding of the parton evolution process.
\par 
The same unintegrated gluon densities have been used to calculate also the 
transverse momentum spectrum of the Higgs boson
production at the Tevatron and the LHC. 
Given the large transverse momenta of the gluons involved, \kt-factorization is
the appropriate tool for calculating higher order corrections. However, the
different unintegrated gluon densities show significant effects at LHC energies,
indicating the need for better experimental constraints as well as further
theoretical studies for a more detailed understanding of parton evolution at
large energies.
\section*{Acknowledgments}
I wish to thank R~Godbole for drawing my interest to Higgs production in the
\kt-factorization approach.
I am grateful to F~Hautmann for explanations of his calculation of Higgs
production in \kt-factorization. I wish to thank J~Gayler and L~Jönnson 
for many comments and a careful reading of the manuscript
and the DESY directorate for hospitality and support..
\section*{References}
\raggedright

\end{document}